\documentclass[12pt,epsfig]{article}
\usepackage{graphicx,amsmath,amssymb}
\def\be{\begin{equation}}
\def\ee{\end{equation}}
\def\bea{\begin{eqnarray}}
\def\eea{\end{eqnarray}}

\begin{document}

\title{\Large \bf Some Problems of Diffraction at High Energies }
\author{\large A. B. Kaidalov \bigskip \\
{\it  Institute of Theoretical and Experimental Physics, Moscow}}
\date{}
\maketitle

\begin{abstract}
An interplay of perturbative and nonperturbative effects in
  pomeron, odderon and reggeons dynamics is discussed.
  It is pointed out that experimental data
  on pion charge exchange reaction at high energies indicate
  to a dominance of the nonperturbative string-like dynamics
  up to rather large   momentum   transfer.
  Role of shadowing effects related to triple -
  pomeron interactions   is investigated.
\end{abstract}

\section{Perturbative versus nonperturbative dynamics for pomeron,
   odderon and reggeons} \label{s1}
Interactions of hadrons at high energies and of virtual photons at
small Bjorken--$x$ are usually described in terms of reggeon
exchanges. For total cross sections and diffractive processes the
pomeron exchange plays the dominant role. It has the vacuum
quantum numbers: signature $\sigma=+$, parity $P=+$ and
C--parity=+. In QCD the pomeron is usually related to gluonic
exchanges in the t--channel. In this talk I shall discuss the
dynamical origin of the pomeron in QCD and a relative role of
perturbative and nonperturbative effects.

Perturbative QCD in the leading logarithmic approximation  leads
to the pomeron singularity with an intercept above unity, which
corresponds to summation of a ladder-type diagrams with exchanges
of reggeized gluons ~\cite{BFKL} -- BFKL pomeron. An increase with
energy of total cross sections is determined by the value of
 $\Delta \equiv \alpha_P(0)-1$: $\sigma^{tot}\sim s^{\Delta}$.
In the leading approximation the BFKL  pomeron has
 $\Delta=\frac{12ln2}{\pi}\alpha_s\approx 0.5$,
which corresponds to a very fast increase of total cross sections
and parton densities with energy ($1/x$). In this approximation
the pomeron corresponds to a cut in j-plane. In the next to
leading approximation  there is a sequence of poles in j-plane
concentrated at point j=1 and the value of $\Delta$ substantially
decreases to the values $\Delta=0.15\div 0.25$ \cite{Fadin}. The
rightmost in j-plane pole strongly depends on nonperturbative
region of small momentum transfer \cite{Kancheli}. Thus, even in
the perturbative approach it is important to have a
nonperturbative input in order to describe asymptotic behavior of
scattering amplitudes.

  The role of nonperturbative effects in pomeron dynamics and a connection
of the pomeron with spectrum of glueballs were studied in refs.
\cite{Simn}, using the method of vacuum correlators  \cite{S90}.
This method, based on general properties of correlators in QCD
 leads to a successful description of the spectrum of resonances for usual
Regge-trajectories and leads to the spectrum of glueballs in a
good agreement with lattice calculations~ \cite{Simn}. It was
shown~ \cite{Simn} that confinement effects and mixing of gluonic
and $q\overline{q}$--Regge-trajectories are important for the
pomeron dynamics. Note, that the intercept of the pomeron in this
approach can be close to the one predicted in the perturbative
QCD. The arising picture of the vacuum trajectories dominated by
nonperturbative effects at large $t > 0$ and by perturbative
dynamics at large negative t, strongly curved in the region $t\sim
0$ is similar to one obtained from gauge/string duality
\cite{Brower}. Thus the pomeron has a rich dynamics in QCD: both
perturbative and nonperturbative effects and mixing with light
quarks are important in the region of not large $t$.

 In perturbative QCD approach there is a singularity in j-plane close to unity
 with quantum numbers:  $\sigma=-$, parity $P=-$ and $C--parity=-$, ~\cite{odd} --"odderon".
 It is due to exchanges in the t--channel by 3 gluons. In the nonperturbative
 approach, discussed above, situation is quite different. There are states
 made of three gluons, but their masses are rather large (more than 3
 GeV) for the lowest states and extrapolations of their Regge
 trajectories to $t=0$
 lead to negative intercepts ~\cite{Simn}. Mixing with $q-\bar q$ states in the
 small--t region in this case is not essential. Thus in this approach there
is no odderon, which can influence asymptotic behavior of
scattering amplitudes.
 Same conclusion has been obtain in ref.~\cite{Teper} from lattice calculations.
 Thus an experimental search for the "odderon" singularity is very important
 as it allows to separate between perturbative and nonperturbative mechanisms
 in QCD at high energies. H1 collaboration has looked for odderon exchange in
 reactions $\gamma p\rightarrow \pi^0N(N^*),~ \gamma p\rightarrow \pi^0\pi^0N(N^*)$,
 which are due to exchanges with $C=-$ in the t--channel and has not found such
 events. The limits on cross sections of these reactions are much lower than
 predictions of ref.~\cite{Dosh}. It is not excluded, however, that couplings of the
 odderon to hadrons are smaller than in the model of ref.~\cite{Dosh}.

 An important information on dynamical properties of reggeons can be obtained from
 a study of $q-\bar q$ Regge--trajectories with isospin I=1. In perturbative QCD
 such singularities have $\alpha > 0$ and tend to zero for $t\to -\infty$ \cite{Kirschner}.
 On the other hand the nonperturbative string--like dynamics leads to linear
 Regge trajectories.

 The leading  $\rho, A_2$ –- trajectories are well determined
 experimentally at  $t > 0$ from the  spectrum of
 resonances and at  $t < 0$ from analysis of the reactions $\pi^-p\rightarrow
 \pi^0N(X),~ \pi^-p\rightarrow\eta^0N(X)$.

    The $\rho$--trajectory, extracted from the data on the
 reaction  $\pi^-p\rightarrow\pi^0N$, is shown in Fig.1. It is
  very close to  the straight
line passing through $\rho$--meson and higher resonances on
$\rho$-- trajectory. It is important that effective
$\rho$--trajectory closely follows the nonperturbative straight
line up to rather large values of $-t\sim 2~ GeV^2$, reaches the
value $\alpha_{\rho}(-2~GeV^2)\approx -1.2$ and does not show any
indication for moving toward the PQCD prediction. This problem was
also studied in ref.~\cite{Brodsky}, using data on inclusive
$\pi^0$ production in $\pi^-p$--collisions. In this case the data
exist up to even higher values of (-t)~\cite{Kennett}. However the
values of $(1-x)$ are not very small ($\sim 0.1$) and PQCD
contribution can not be excluded~\cite{Brodsky}. In the exclusive
charge exchange reaction, analyzed above, the values of $s\sim
10^3$ and limits on PQCD contribution are more stringent (the
ratio of couplings to hadrons for PQCD and string--like
contributions is $< 10^{-3}$).

\begin{figure}
\vskip0.05in \centering
\includegraphics[width=3in]{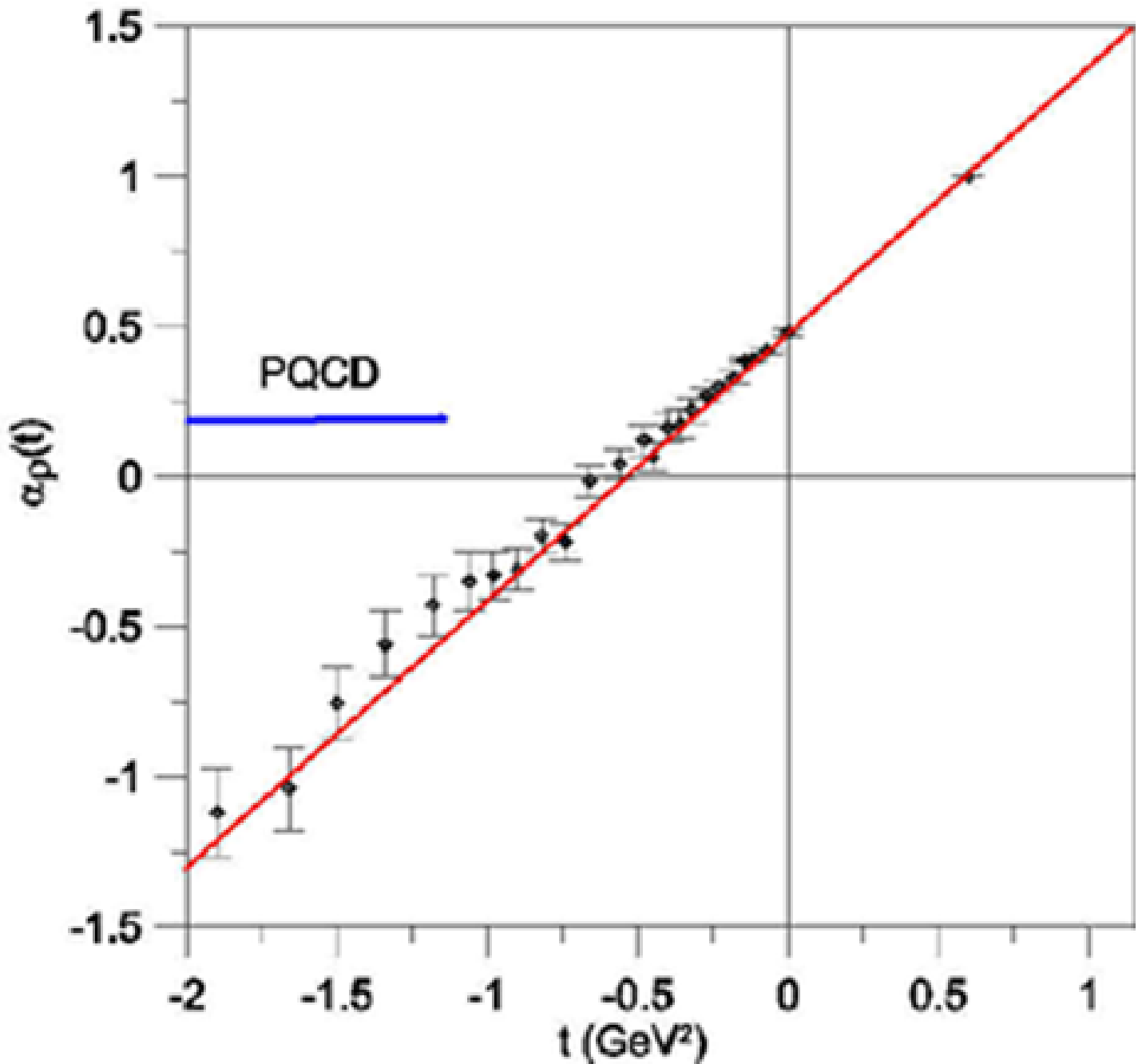}
\caption{ Effective $\rho$-trajectory and PQCD preiction.}
\label{fig:rho-trajectory}
\end{figure}

Thus I came to the conclusion that in the high-energy Regge limit
nonperturbative effects play an important role up to rather large
values of $(-t)$.

\section{ Unitarity effects for hard diffractive processes and the role of
 triple-pomeron interactions.}

The pomeron pole with $\Delta > 0$ leads for $s\equiv
W^2\rightarrow \infty$ to a violation of s-channel unitarity. It
is well known that the unitarity is restored if the multi--pomeron
exchanges in the t-channel are taken into account.

An important role of multi--pomeron exchanges is clearly seen in
the hard diffractive processes. In these processes they lead to a
violation of both Regge and QCD factorization. Diffractive
production of jets, W-bosons, heavy quarks and heavy quarkonia was
observed by CDF and $D_0$--collaborations at Tevatron. There are
hopes to study Higgs bosons in exclusive double diffractive
production at LHC (see for example~\cite{Martin}).

For diffractive dijet production at Tevatron it was found that
calculation of the cross section, based on the factorized formula
with diffractive structure functions
 from HERA data, leads to a large discrepancy with the CDF measurements
 both in the normalization and in the shape of the observed distribution.
 This factor of $\sim 10$ difference between prediction of factorization and
 experiment is naturally accomodated by suppression due to multi--pomeron
 exchanges. The Gribov reggeon diagrams technique~\cite{regtech} allows one
to calculate contributions of these exchanges (Regge cuts) to
scattering amplitudes. The diagrams with n--pomeron exchange in
the t--channel can be expressed as sums over all possible
intermediate diffractive states in the s--channel. The sum of the
elastic rescatterings leads to the eikonal formula. Inclusion of
diffraction dissociation to not large masses leads to a
multichannel generalization of the eikonal approximation.
 The two-channel model for calculation of suppression (survival probability)
 for hard diffractive processes has been used in ref.~\cite{KKMR1} and lead to
 a reasonable description of experimental data on diffractive dijet production
 at Tevatron.

 Diffractive production of large mass states is described in reggeon approach
by diagrams with interactions between pomerons. For hard processes
it leads to a new class of multi--pomeron diagrams, indicated for
dijet production on Fig.2. This interaction can happen if the mass
of the hadronic state above the hard part of Fig.2 a) is very
large and rapidity intervals in Fig2 b) $y_1, y_2 \gg 1$. It was
pointed out in ref. \cite{KKMR1} that this condition is not
satisfied for production of dijets at Tevatron and of Higgs at LHC
and they are not essential for calculation of survival
probabilities in these cases (see also discussion in
ref.~\cite{Khoze}). Investigation of neutron spectra in
photoproduction at HERA`in the reggeized pion exchange model
~\cite{neutrons} also indicates that interactions of pomerons
 are not very essential in these processes.

\begin{figure}
\vskip0.05in
\begin{center}
\includegraphics[width=0.5\textwidth]{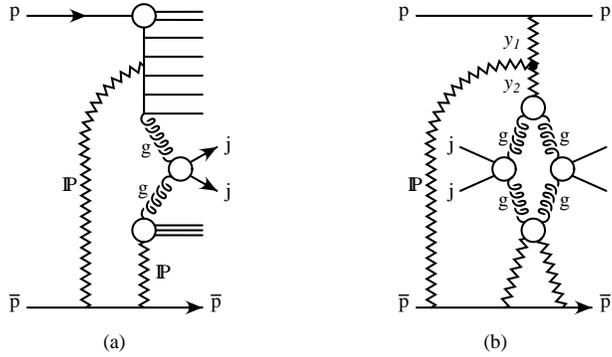}
\caption{ Diagrams with pomeron interactions for hard
diffraction.}
\label{fig:2}
\end{center}
\end{figure}

However in general at superhigh energies and very large masses of
diffractively produced hadronic states these diagrams can be
important. This problem has been emphasized by J.~Bartels et
al.~\cite{Bartels} in QCD perturbation theory.

It may be instructive to consider the problem of influence of
pomeron interactions in a simple and  solvable model. For this
purpose I shall use the Schwimmer model~\cite{Schwimmer}. It
corresponds to summation of fan type diagrams with triple pomeron
interactions. the model can give reasonable approximation for
amplitudes when the size of a projectile is much less than the
size of a target. Diffractive cross section for production of a
state of a given mass $M$ and rapidity gap $y$ (with the value of
$y_M = Y - y$ fixed; Y=$lns/s_0$) in this model is
known~\cite{Levin,Boreskov} and has the form
\begin{equation}
\label{eq:sig_M_sch} \frac{d\sigma^{\rm D}}{dy_M} \equiv
M^2\frac{d\sigma^{\rm D}}{dM^2}= - g_1 g_2 \frac{d\varphi_{\rm
gap}(Y;y)}{dy} = \frac{2 g_1 g_2\Delta\epsilon\,e^{\Delta (2 Y -
y_M)}}{[1+\epsilon\,(2e^{\Delta Y}-e^{\Delta (Y - y_M)}-1)]^2}
\end{equation}
 with $g_1, g_2$ -couplings of the pomeron with particles 1 (small size) and 2
 (large size) correspondingly; $\epsilon = \frac{r g_2}{\Delta}$,
where $r$-is the triple pomeron coupling.
   Thus in this model the survival probability is:
\be
 S^2_{Sch}=\frac{1}{[1+\epsilon\,(2e^{\Delta Y}-e^{\Delta (Y - y_M)}-1)]^2}
\ee Contrary to the eikonal model it depends not only on $Y$, but
also on $y (y_M$). As expected, interactions of pomerons lead to
suppression of diffractive cross sections calculated in the pole
approximation.

These results can be generalized to the eikonalized Schwimmer
model \cite{Boreskov}:
\begin{equation}
\label{surv_prob}
 S^2=S^2_{eik}S^2_{Sch}
\end{equation}
where
 \be \label{surv_prob} S^2_{eik}(Y,y_M;b)= \exp\left[-g_1 g_2
(\varphi_{tot}(Y;b) - \varphi_{gap}(Y,Y-y_M;b))\right] ~. \ee and
\begin{equation}
\varphi_{tot}(Y) = \frac{2 P(Y)}{1+\epsilon [P(Y)-P(0)]},~~ P(Y) =
{\rm exp}(\Delta Y).
 \label{eq:sol}
\end{equation}

It is important that due to pomeron interactions the value of the
eikonal function is reduced and it does not increase for $s\to
\infty$. This leads to a strong increase of survival probability
compared to the pure eikonal approximation. Thus there is an extra
decrease of $S^2$ due to $S^2_{Sch}$ term and its increase due to
change of $S^2_{eik}$. For very large $Y$ the second effect is
more important and inclusion of pomeron interactions leads in this
model not to a decrease, but to an increase of survival
probability compared to the pure eikonal approximation.

Let us note that in any case cross sections of inelastic
diffraction are very small for small impact parameters at very
high energies and the main problem is to calculate these cross
sections at large impact parameters (the edge of interaction
region). In this region nonperturbative effects are important and
any calculation on PQCD is nor reliable.

Thus it is important to develop a reliable nonperturbative method
of calculation in QCD of absorptive effects in diffractive
processes with account of pomeron interactions (including pomeron
loops).

\section{Acknowledgments}

 I would like to thank K.Boreskov, O.V. Kancheli, V.A. Khoze, L.N. Lipatov, A.
 Martin, M.G. Ryskin for useful discussions.
 This work is supported in part by the grants: CRDF RUP2-2621-MO-04, RFBR 04-02-17263, Science
  Schools 1774.2003. \\


\end{document}